\documentclass[preprint]{elsarticle}

\usepackage{lineno,hyperref}
\usepackage{multicol}
\usepackage{multirow}
\usepackage{graphicx}
\usepackage{booktabs}
\usepackage{color}
\usepackage{geometry}
\begin{document}

\begin{frontmatter}

\title{3-D topological signatures and a new discrimination method for single-electron events and 0$\nu$$\beta$$\beta$ events in CdZnTe: A Monte Carlo simulation study\tnoteref{mytitlenote}}
\tnotetext[mytitlenote]{Supported by National Natural Science Foundation of China (11305093 \& 11175099) and Tsinghua University Initiative Scientific Research Program (2011Z07131 \& 2014Z21016)}

\author[mymainaddress,mysecondaryaddress]{Ming Zeng}
\author[mymainaddress,mysecondaryaddress]{Teng-Lin Li}
\author[mymainaddress,mysecondaryaddress]{Ji-Rong Cang}
\author[mymainaddress,mysecondaryaddress]{Zhi Zeng\corref{mycorrespondingauthor}}
\cortext[mycorrespondingauthor]{Corresponding author}
\ead{zengzhi@tsinghua.edu.cn}
\author[mymainaddress,mysecondaryaddress]{Jian-Qiang Fu}
\author[mymainaddress,mysecondaryaddress]{Wei-He Zeng}
\author[mymainaddress,mysecondaryaddress]{Jian-Ping Cheng}
\author[mymainaddress,mysecondaryaddress]{Hao Ma}
\author[mymainaddress,mysecondaryaddress]{Yi-Nong Liu}

\address[mymainaddress]{Key Laboratory of Particle \& Radiation Imaging (Tsinghua University), Ministry of Education, China}
\address[mysecondaryaddress]{Department of Engineering Physics, Tsinghua University, Beijing 100084, China}

\begin{abstract}
In neutrinoless double beta (0$\nu$$\beta$$\beta$) decay experiments, the diversity of topological signatures of different particles provides an important tool to distinguish double beta events from background events and reduce background rates. Aiming at suppressing the single-electron backgrounds which are most challenging, several groups have established Monte Carlo simulation packages to study the topological characteristics of single-electron events and 0$\nu$$\beta$$\beta$ events and develop methods to differentiate them. In this paper, applying the knowledge of graph theory, a new topological signature called \textit{REF track} (Refined Energy-Filtered track) is proposed and {proven} to be an accurate approximation of the real particle trajectory. Based on the analysis of the energy depositions along the \textit{REF track} of single-electron events and 0$\nu$$\beta$$\beta$ events, the \textit{REF energy deposition models} for both events are proposed to indicate the significant differences between them. With these {differences}, this paper presents a new discrimination method, which, in the Monte Carlo simulation, achieved a single-electron rejection factor of 93.8 $\pm$ 0.3 (stat.)\% as well as a 0$\nu$$\beta$$\beta$ efficiency of 85.6 $\pm$ 0.4 (stat.)\% with optimized parameters in CdZnTe.
\end{abstract}

\begin{keyword}
0$\nu$$\beta$$\beta$\sep CdZnTe\sep topological signature \sep \textit{REF track} \sep background reduction
\end{keyword}
\end{frontmatter}

\section{Introduction}
Research on neutrinoless double beta decay is a key approach to answer whether neutrinos are Majorana particles and whether the total lepton number is conserved in nature; in addition, 0$\nu$$\beta$$\beta$ provides direct information about the absolute neutrino mass scale \cite{lab1}. Several experiments using different 0$\nu$$\beta$$\beta$ isotopes are already running (or are about to begin) to explore the best strategy for 0$\nu$$\beta$$\beta$ searches, for example, the $^{76}$Ge-based GERDA experiment \cite{lab2}, the $^{136}$Xe-based EXO-200 experiment \cite{lab3}, the $^{130}$Te-based CUORE experiment \cite{lab4}, {the $^{76}$Ge-based MAJORANA experiment} \cite{lab23} and {the $^{136}$Xe-based PandaX-III experiment} \cite{lab22}. One formidable challenge of these experimental approaches is their capability to control the backgrounds to the required extremely low levels \cite{lab1}. {To achieve more powerful means of background identification, the topology concepts \cite{lab21} and the graph theory \cite{lab17} were introduced in 0$\nu$$\beta$$\beta$ decay experiments for further background suppressions.} Among all the backgrounds, the single-electron events are the most challenging ones due to their similar behaviours compared with two-electron signals, which draw attention of several large cooperations:

The COBRA experiment \cite{lab5} is planning to use a large amount of CdZnTe room temperature semiconductor detectors for the search for 0$\nu$$\beta$$\beta$. CdZnTe contains 9 candidate isotopes for 0$\nu$$\beta$$\beta$, and among them, $^{116}$Cd has the highest \textit{Q}-value of 2813.50 keV. This isotope is the main target of the search because its \textit{Q}-value lies above the highest relevant naturally occurring gamma line ($^{208}$Tl at 2614.6 keV), thereby contributing to the background reduction in the region of interest. Another possible means to reach the extremely low background level \cite{lab6} could be the use of pixelated detectors to exploit the topological information of the events. In Ref \cite{lab7}, the COBRA collaboration developed a set of selection criteria between single-electron and 0$\nu$$\beta$$\beta$ events in pixelated CdZnTe detectors based on three distributions of the 2-D topological signatures: the number of active pixels, the maximum separation of active pixels, and the variation in energy depositions{. With the criteria, a discrimination efficiency of 70\% for both single-electron and 0$\nu$$\beta$$\beta$ events was achieved.} However, {the three distributions} are proposed based on the features of 2-D unorganized pixels, and the 3-D topological signatures of single-electron and 0$\nu$$\beta$$\beta$ events in CdZnTe have not been discussed yet.

The NEXT experiment \cite{lab8} is planning to search for 0$\nu$$\beta$$\beta$ in high pressure xenon gas TPC. One main advantage of this technique is the ability to reconstruct the trajectory of the two electrons emitted in the decays, which contributes to background suppression. In Ref. \cite{lab17}, the NEXT collaboration developed an inspiring algorithm to group the voxels into tracks and find their end-points and proposed a method to distinguish single electrons and electron-positron pairs based on the energy deposited at the end-points of the track (called a `blob'){. With the method, they achieved a 0$\nu$$\beta$$\beta$ signal efficiency of 40\% and reduced the background (single-electron events and others) level by around three orders of magnitude.} However, the topological signature of the `lower energy blob candidate' does not take full advantage of the entire track and lacks the robustness as a discrimination method.

\begin{figure}[!htb]
    \begin{center}
    \includegraphics[width=15cm]{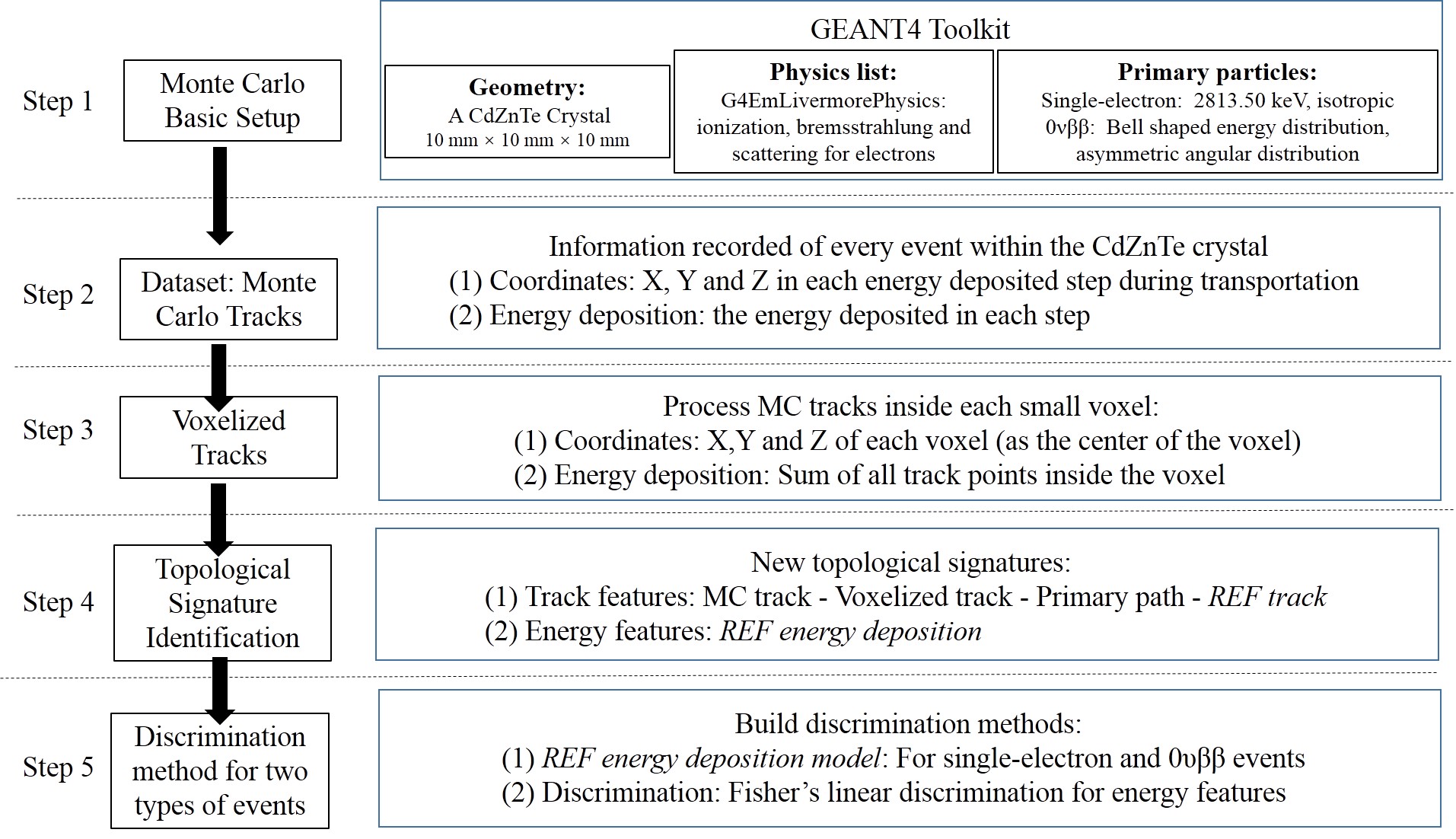}
    \caption{\label{fig1} {Structure of the work presented in this paper. The key points in each step are listed.} Step 1, 2, and 3 are introduced in section 2. Step 4 {is} introduced in section 3. Step 5 {is} introduced in section 4.}
    \end{center}
\end{figure}

In this paper, two new topological signatures about track and energy features are proposed for the 3-D trajectories of single-electron and 0$\nu$$\beta$$\beta$ events in the Monte Carlo simulation. Based on the energy topological features, a new discrimination method is proposed to differentiate two types of events. The whole process of the method is shown in Fig.~\ref{fig1} and section 2-4. The discussions are made in section 5. The conclusion is summarized in section 6.

\section{The Monte Carlo simulation method}

\subsection{The basic setup for simulation}
{To simulate of the behaviour of single-electron events and 0$\nu$$\beta$$\beta$ events, the GEANT4 simulation package(Version: 4.10.2) \cite{lab10} is used in this paper. The G4EmLivermorePhysics in Low Energy electromagnetic physics lists \cite{lab11} is adopted, which combines Livermore with Standard Electromagnetic Physics models and performs better in lower energy. The corresponding physical processes affecting electrons are ionization, bremsstrahlung, and scattering. The cut-off for gamma, electron and positron is set as 1$\mu$m, which is accurate enough compared with the 0.1 mm - voxelization introduced in section 2.3.

CdZnTe is chosen as the object of study in this paper, which is one of the typical materials in the 0$\nu$$\beta$$\beta$ study. Note that the analysis of topological signatures and discrimination methods introduced in this paper would be appropriate for other materials as well. In CdZnTe, the interested 0$\nu$$\beta$$\beta$ elements is $^{116}$Cd with the highest \textit{Q}-value of 2813.50 keV. For a simplified study of the topological signatures, the single-electron (0$\nu$$\beta$$\beta$) events are assumed to occur at the central point of CdZnTe and produce one (two) emitted electron(s). {It is roughly equivalent compared with generating events at random positions and applying a near-edge cut.} The volume of CdZnTe is 10 mm $\times$ 10 mm $\times$ 10 mm, which is able to hold most electrons' track and stop parts of the bremsstrahlung photons if the electrons undergo bremsstrahlung process. A pre-selection is made to exclude the events that did not deposit all their energy in CdZnTe (mainly due to the escape of bremsstrahlung photons). For a 10 mm $\times$ 10 mm $\times$ 10 mm CdZnTe, about 74\% single-electron events and 81\% 0vbb events will pass the pre-selection.

In the 0$\nu$$\beta$$\beta$ decay of $^{116}$Cd, two electrons are emitted simultaneously, with a total energy of 2813.50 keV. As the mechanism of 0$\nu$$\beta$$\beta$ decay is not well understood, there are many hypotheses providing different energy and angular distributions of the two electrons, as discussed in Ref. \cite{lab12}. The simulation program in this paper considered the \textit{mass} \textit{mechanism} as the domain mechanism, to remain consistent with the choice in Ref. \cite{lab7}. In this 0$\nu$$\beta$$\beta$ hypothesis, the energy distribution of each electron is bell shaped and centred on half of the \textit{Q}-value. The angular distribution between the two electrons is asymmetric shaped, and the peaks are at the opening angle of $\theta$ = 120$^\circ$. The curves of both distributions can be found in Ref. \cite{lab7}.

The simulation program also generated single-electron events to discuss the differences of their topological signatures from 0$\nu$$\beta$$\beta$ events. The single-electron is emitted from the central point of CdZnTe in a random direction. The initial energy of the electron is 2813.50 keV, corresponding to the \textit{Q}-value of $^{116}$Cd. In a real CdZnTe detector system, the single-electron background probably origins from the $\beta$-decay of $^{214}$Bi (\textit{Q} = 3.272 MeV) or a higher-energy photon produced by a excited nuclei (from neutron capture reactions, probably).
}

\subsection{Monte Carlo tracks}
\begin{figure}[!htb]
    \begin{center}
    \includegraphics[width=15cm]{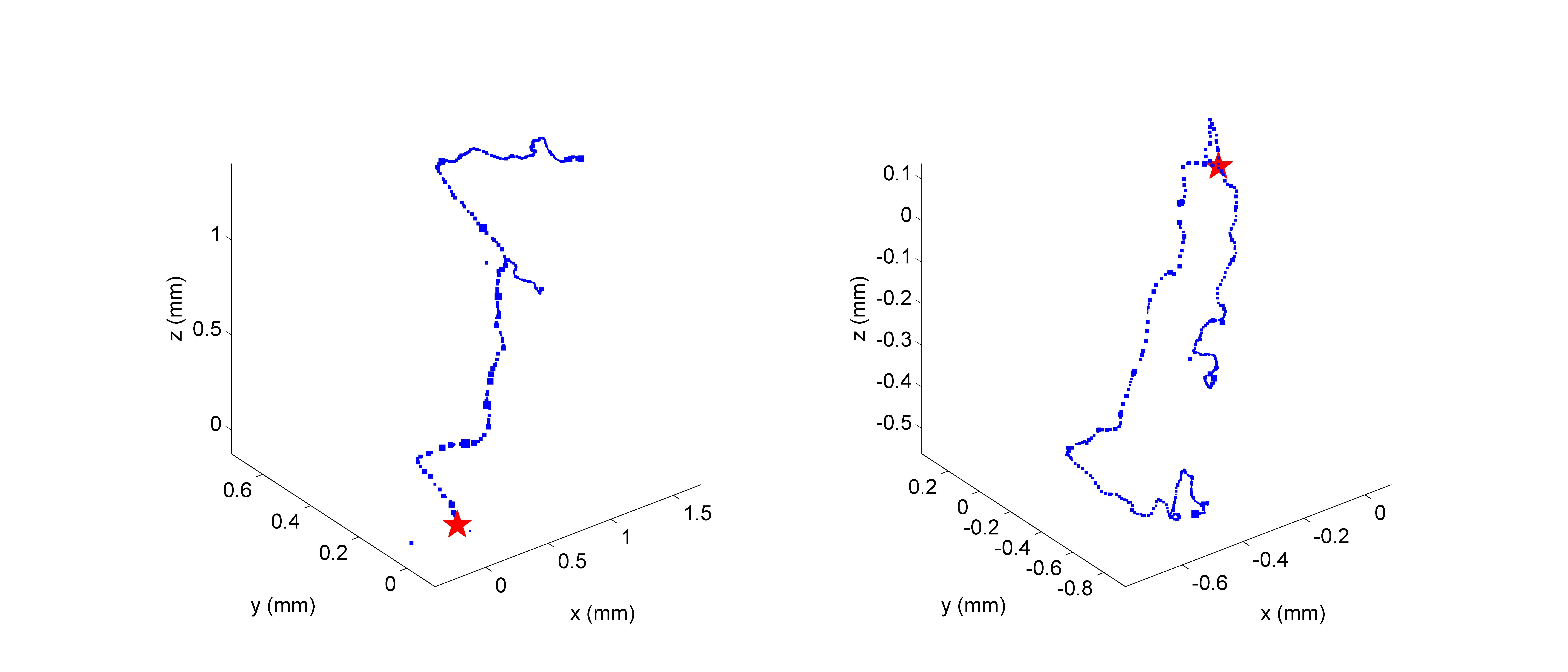}
    \caption{\label{fig2} An instance of the MC tracks of a single-electron event (left) and a 0$\nu$$\beta$$\beta$ event (right). It is difficult to differentiate them with naked eye. {The star indicates the starting point (0, 0, 0) of both events.} }

    \includegraphics[width=15cm]{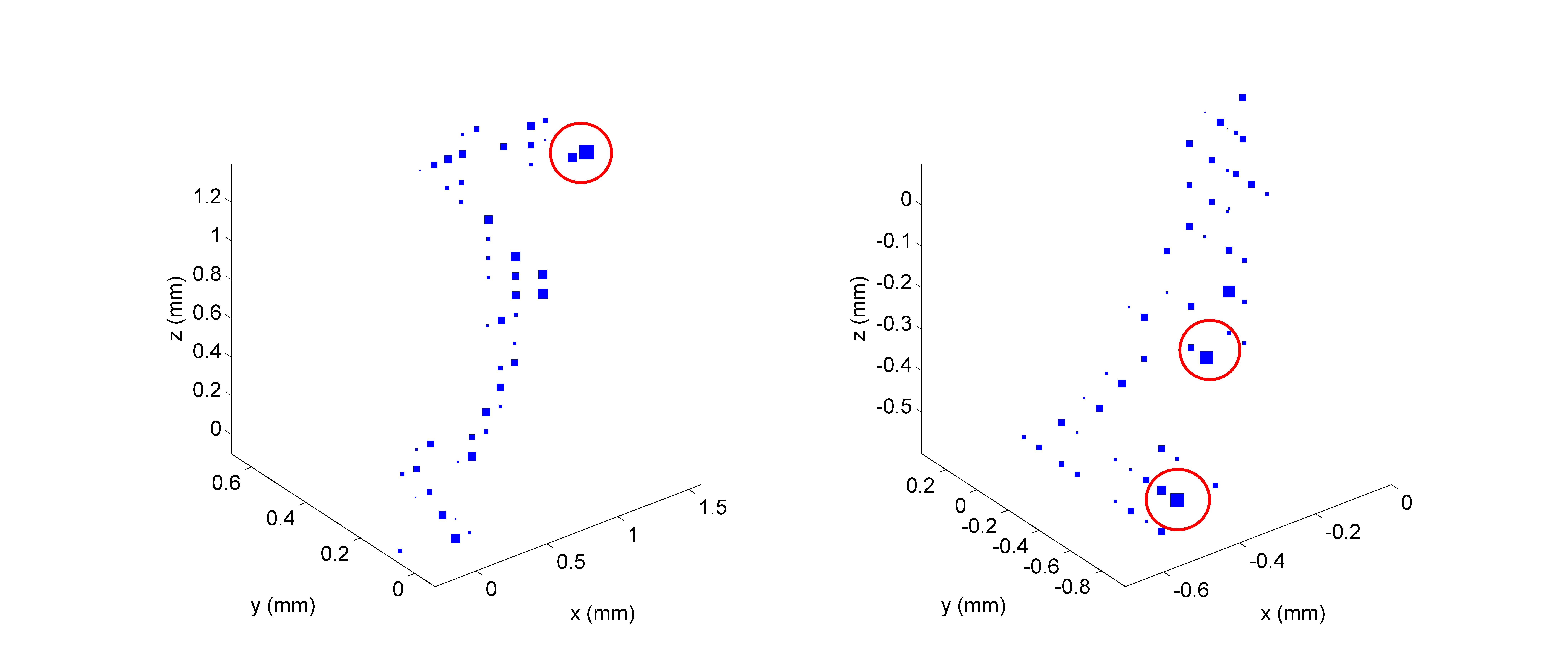}
    \caption{\label{fig3} An instance of the voxelized tracks of a single-electron event (left) and a 0$\nu$$\beta$$\beta$ event. {The marker size is proportional to the energy deposition (similarly hereafter). The largest marker here is about 0.3 MeV.} The energy depositions at the endpoints are characterised by: one larger energy deposition ({marked by one circle}) at one endpoint of a single-electron event and two larger energy depositions ({marked by two circles}) at both endpoints of a 0$\nu$$\beta$$\beta$ event.}
    \end{center}
\end{figure}

For both {kinds of} single-electron events and 0$\nu$$\beta$$\beta$ events generated by GEANT4, the x, y, z-coordinates and energy deposited at each step are recorded in the dataset. This track of a single-electron event or 0$\nu$$\beta$$\beta$ event with energy-deposition information is called a `Monte Carlo (MC) track'. An instance of the MC tracks for both events is shown in Fig.~\ref{fig2}.

\subsection{Voxelization and voxelized track}
To study the topological signatures of the single-electron and 0$\nu$$\beta$$\beta$ events in a practical situation, the MC tracks are voxelized to a limited spatial resolution. The initial voxel {is a 0.1 mm $\times$ 0.1 mm $\times$ 0.1 mm cube.} (\textit{D} = 0.1 mm, \textit{D} is the side length of the voxel), which represents a 3-D spatial resolution of 0.1 mm. {This might be technologically achievable \cite{lab13}}. Furthermore, worse spatial resolutions (\textit{D} = 0.2 mm to 0.5 mm ) are discussed in section 5.3 for degraded situations.

The process of voxelization is to calculate a voxelized track from the MC track: The coordinate of each point in the voxelized track is given by the central coordinate of the voxel, and the energy of each point is given by the sum of the energy deposited within the voxel's boundaries. An instance of the voxelized tracks is shown in Fig.~\ref{fig3}, which are the voxelization of the MC tracks in Fig.~\ref{fig2}.

\subsection{Spatial end energy distribution of the voxelized track}
\begin{figure}[!htb]
    \begin{center}
    \includegraphics[width=15cm]{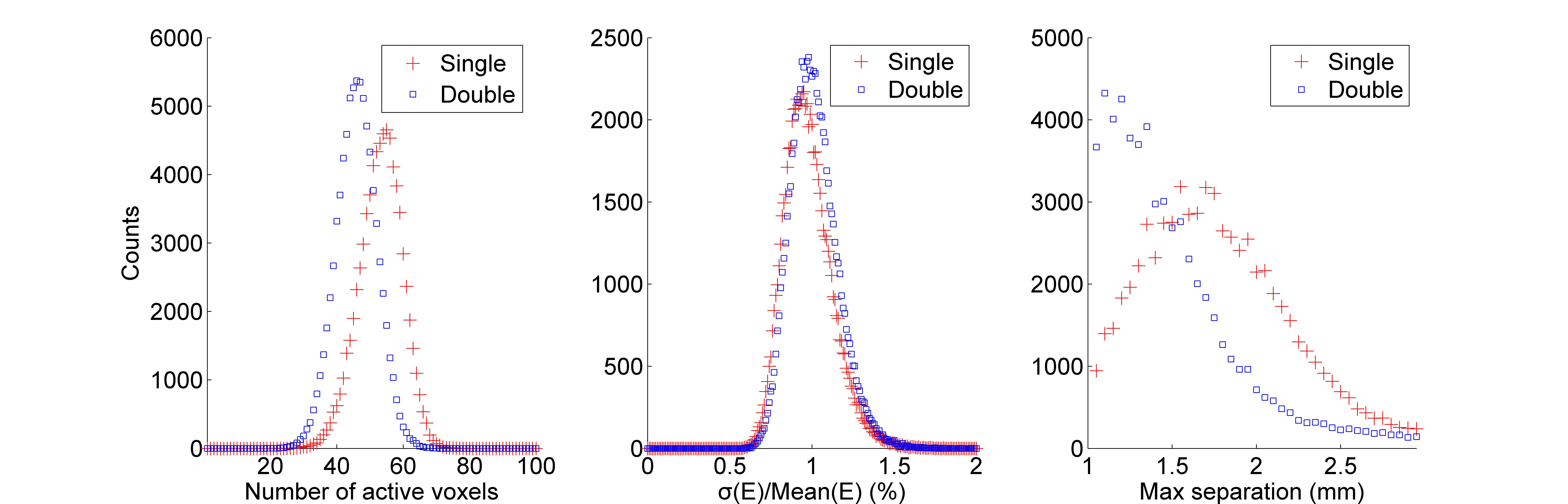}
    \caption{\label{fig4} {Signature distributions for} the single-electron events and 0$\nu$$\beta$$\beta$ events for a voxel size of 0.1 mm and an energy of 2813.50 keV. (Left) The number of voxels {with an energy deposition.} (Centre) The standard deviation of the energy divided by the mean energy (in percent) {for individual voxels}. (Right) The maximum separation of any two voxels in the voxelized track. {Differences between events of different types are observed.}}
    \end{center}
\end{figure}

With the voxelized track, several simple 3-D topological signatures can be {defined}: the number of active voxels, the variation in the energy deposition of every voxel, and the maximum separation of active voxels, where an `active' voxel means a voxel with a non-zero energy deposition. The distributions of these 3-D topological signatures is shown in Fig.~\ref{fig4}, and similar to the results introduced by the COBRA collaboration for simple 2-D topological signatures in CdZnTe \cite{lab7}. {We have tried to use these signatures to develop a discrimination method with Support Vector Machine, but ended up with a poor discrimination efficiency. Then we started to find new topological signatures.}

\section{New topological signatures: \textit{REF track} and \textit{REF energy deposition}}

\subsection{Features of the energy deposition in CdZnTe}

\subsubsection{Specific energy loss of the electron in CdZnTe}
\begin{figure}[!htb]
    \begin{center}
    \includegraphics[width=7.5cm]{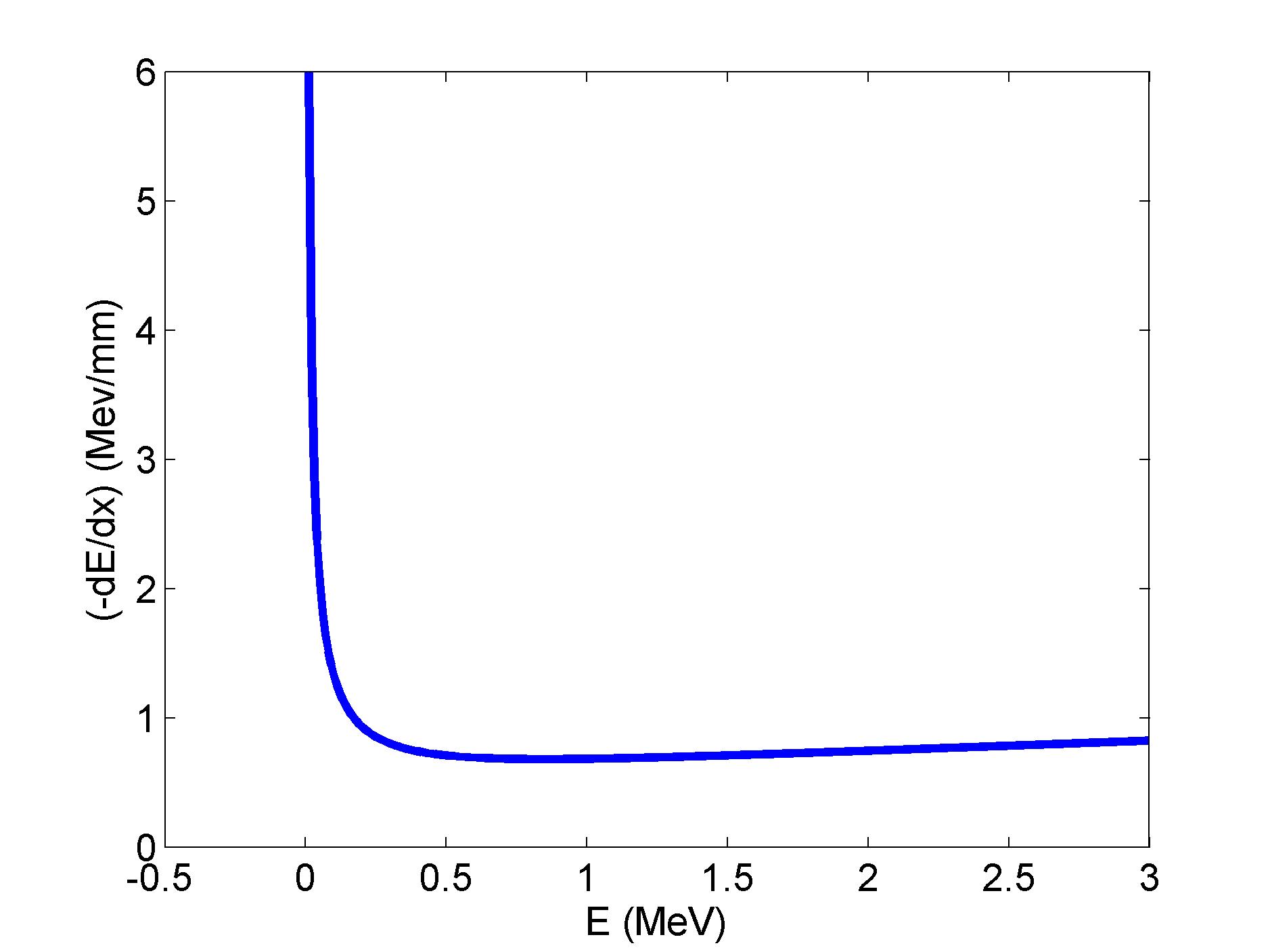}
    \caption{\label{fig5} Specific energy loss of the electron in CdZnTe.}
    \end{center}
\end{figure}

{When traversing matter, electrons lose energy through scattering and radiation \cite{lab14}.} According to the relevant material parameters of CdZnTe, the specific energy loss of electrons in CdZnTe is shown in Fig.~\ref{fig5}.

{The resulting lengths of electron tracks, \textit{L}, depend on their initial energy \textit{E}$_{0}$ as}
\begin{eqnarray}
    \label{eq1}
    L = \int\nolimits_{0}^{\;E_{0}} \frac{1}{-({\rm d}E/{\rm d}x)} {\rm d}E.
\end{eqnarray}

{Note that \textit{L} is not the \textit{electron} \textit{range} and that there are large event-by-event fluctuations. On average, the length of the track of a 2813.50 keV electron is 3.7 mm. (For comparison, the \textit{electron} \textit{range} of a 2813.50 keV electron in CdZnTe is approximately 1.5 mm \cite{lab15}.)}

\subsubsection{Energy deposition at the endpoints}
Fig.~\ref{fig5} shows that the specific energy loss of the electron is approximately constant as the energy of the electron decreases along the length of track, until it reaches the end of the track, where a dramatic increase of energy loss occurs. This increase causes a significant energy deposition in a compact region (so-called `blobs' \cite{lab17}), as shown in Fig.~\ref{fig3}. Another important reason for the occurrence of blobs in voxelized tracks is that a slower electron at the end would meet stronger deflections due to the multiple scattering and thus wrap around in a compact region.

In Fig.~\ref{fig2}, the `blobs' cannot be easily found at the endpoints of the MC tracks due to the impact of the large energy fluctuation, which can be reduced after voxelization. As a result, in Fig.~\ref{fig3}, the signature of the `blobs' are much clearer: One larger energy deposition at one of the endpoints of the voxelized track can be found in the single-electron event. In contrast, the voxelized track of the 0$\nu$$\beta$$\beta$ event is characterized by two larger energy depositions at both endpoints of the voxelized track.

\subsection{Reconstruction of the \textit{REF track}}
In a 3-D position sensitive detector, voxelized tracks comprised of 3-D energy depositions can be measured, and the size of voxels is corresponding to the 3-D spatial resolution of the detecting system. These unorganized spatial energy depositions carry the track and energy features of the particles and need to be interpreted. In this paper, a new track topological signature called \textit{REF track} (Refined Energy-Filtered track) and a new energy topological signature called \textit{REF energy deposition} (energy deposition along \textit{REF track}) are introduced to indicate the characteristics of different particles.

Here are the steps to reconstruct the \textit{REF track}:

The first step is to group the unorganized \textit{points} into a \textit{graph}\cite{lab17}. In the \textit{graph} \textit{theory} in the mathematical field of topology, the \textit{points} are mathematical abstractions for the interconnected objects and the \textit{lines} describe the connectivity between the \textit{points}. In this paper, \textit{points} are defined as the voxelized points (in Fig.~\ref{fig3}) and two \textit{points} are connected only when they are close enough. Thus, a \textit{connectivity criterion} \textit{T} is defined; when the space distance between two \textit{points} is no greater than \textit{\textit{T}}, the \textit{points} are connected with a \textit{line} and the \textit{weight} of the \textit{line} is defined as the space distance.

\begin{figure}[!htb]
    \begin{center}
    \includegraphics[width=15cm]{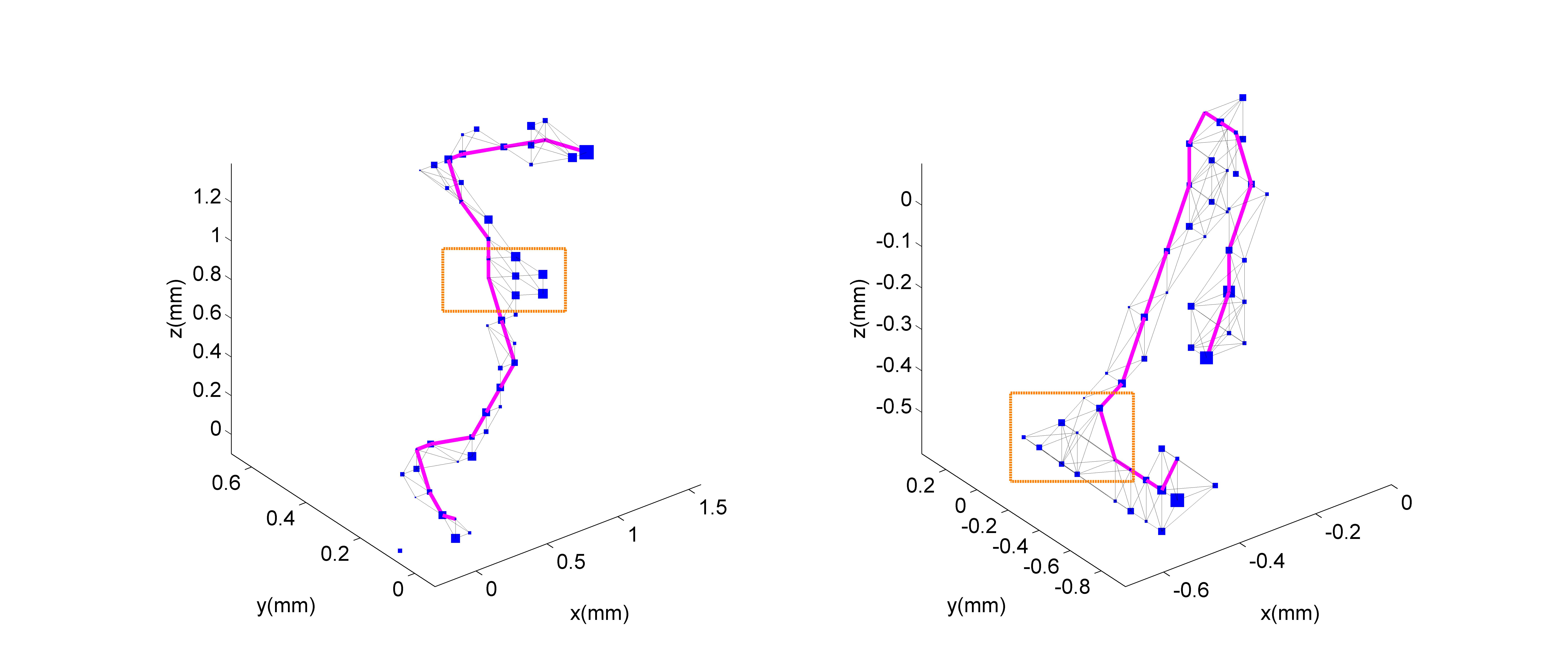}
    \caption{\label{fig6} An instance of the graph (\textit{T} = $\sqrt{3}$\textit{D}) of a single-electron event (left) and a 0$\nu$$\beta$$\beta$ event (right). {Every two \textit{points} with a distance no greater than \textit{T} are connected with a thin line.} The reconstructed \textit{primary path} is marked with the thick line, which fails to mimic the average position of the voxelized track in the dashed rectangle.}

    \includegraphics[width=15cm]{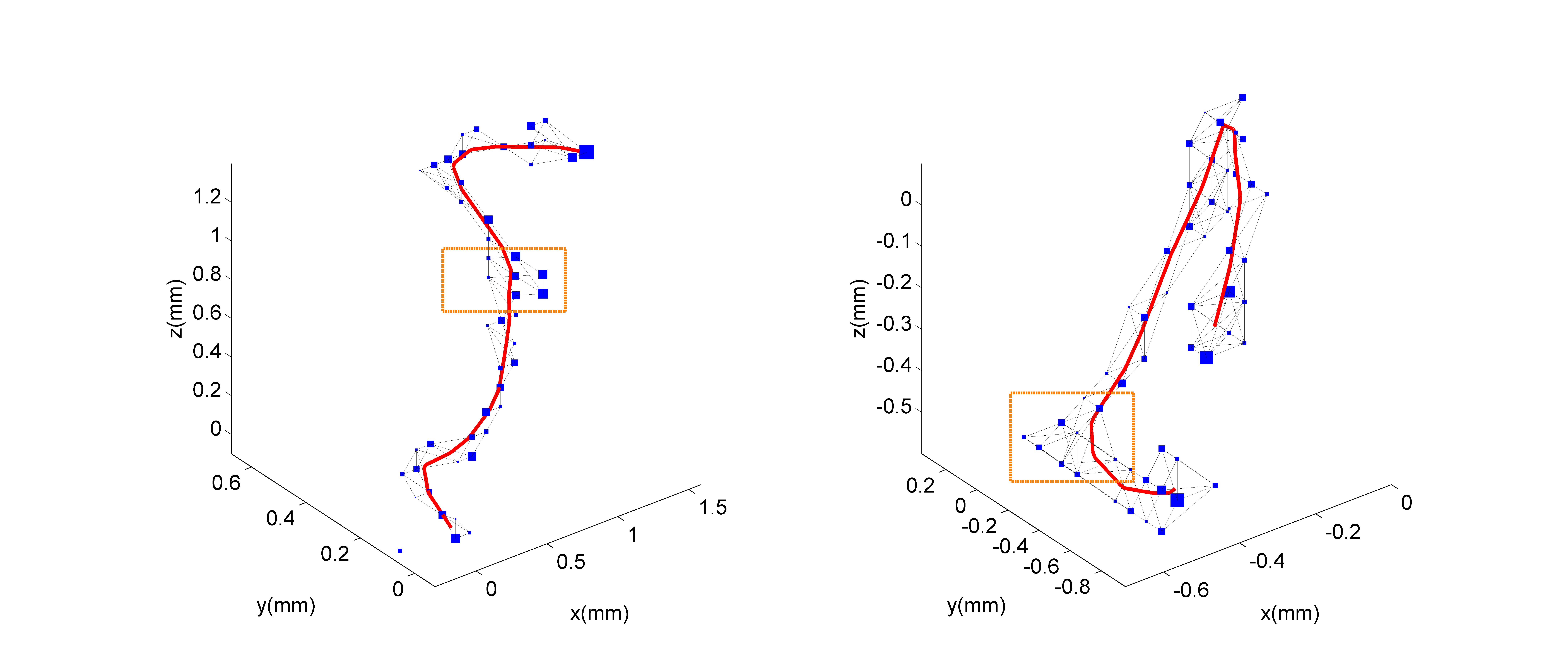}
    \caption{\label{fig7} An instance of the reconstructed \textit{REF track} (thick line) of a single-electron event (left) and a 0$\nu$$\beta$$\beta$ event (right). The \textit{REF track} is smoother than the \textit{primary path} and performs better in the dashed rectangle.}
    \end{center}
\end{figure}

As the \textit{point}'s {3-D} space coordinate is voxelized as a whole-number multiple of \textit{D}, so the {3-D} space distance between \textit{points} can only be $\sqrt{N}\cdot$\textit{D} (\textit{N} as positive integers, \textit{D} as the side length of the voxel). {An instance of the graph with a \textit{connectivity criterion} \textit{T} = $\sqrt{3}$\textit{D} is shown in Fig.~\ref{fig6}. Here "\textit{T} = $\sqrt{3}$\textit{D}" means that two \textit{points} are connected only when the space distance between them is no greater than $\sqrt{3}$\textit{D}.}

According to the typical \textit{shortest} \textit{path} \textit{problems} in \textit{graph} \textit{theory} \cite{lab16}, this algorithm calculates the shortest-path between every two \textit{points} and select the longest of such paths, which is defined as the \textit{primary path}. If multiple paths are found to be the longest paths, then the path with the minimum deflection is selected. {If {an} event deposits its energy into several disconnected tracks in the crystal (by a bremsstrahlung emission, for example), this algorithm will automatically calculate the \textit{primary path} candidate of each disconnected track and select the longest one as the final \textit{primary path}.} An instance of the \textit{primary path} ({thick line}) is shown in Fig.~\ref{fig6}, which can serve as a rough approximation of the trend of the particle's track.

There are also other algorithms to establish a reasonable path, similar to the \textit{primary path}, in order to represent the trend of a real track and identify the end-points. For example, in Ref. \cite{lab17}, {the paths between large-energy-deposition points are calculated and the lengths of them are compared; then the longest path is selected.}

In this paper, the \textit{REF track} (Refined Energy-Filtered track) is proposed as a more reasonable track topological signature. As mentioned above, the \textit{primary path} is a deduction from the calculation of \textit{graph theory}, which doesn't consider the energy topology of an electron's physical track and sometimes fails to mimic the average position of the voxelized track. To improve the \textit{primary path}, a spatial energy filter is established for adjusting this polygonal line according to the energy deposition information: every \textit{point} \textit{P}$_{k}$ on the \textit{primary path} is replaced by the centre-of-gravity \textit{J}$_{k}$ of the energy deposition in an \textit{R}-radius ball around \textit{P}$_{k}$. All of the \textit{J}$_{k}$ constitute a new line, defined as a \textit{REF track} (denoted by the red line in Fig.~\ref{fig7}). The radius \textit{R} of the ball affects the details of the \textit{REF track} and is initially set to $\sqrt{3}$\textit{D}. The influence of the value of \textit{R} is discussed in section 5.2. The \textit{REF track} is a significant improvement compared with the \textit{primary path}, which is graphically shown in the {dashed} rectangle in Fig.~\ref{fig6} and Fig.~\ref{fig7} and supported by specific data in section 5.2.

\begin{figure}[!htb]
    \begin{center}
    \includegraphics[width=15cm]{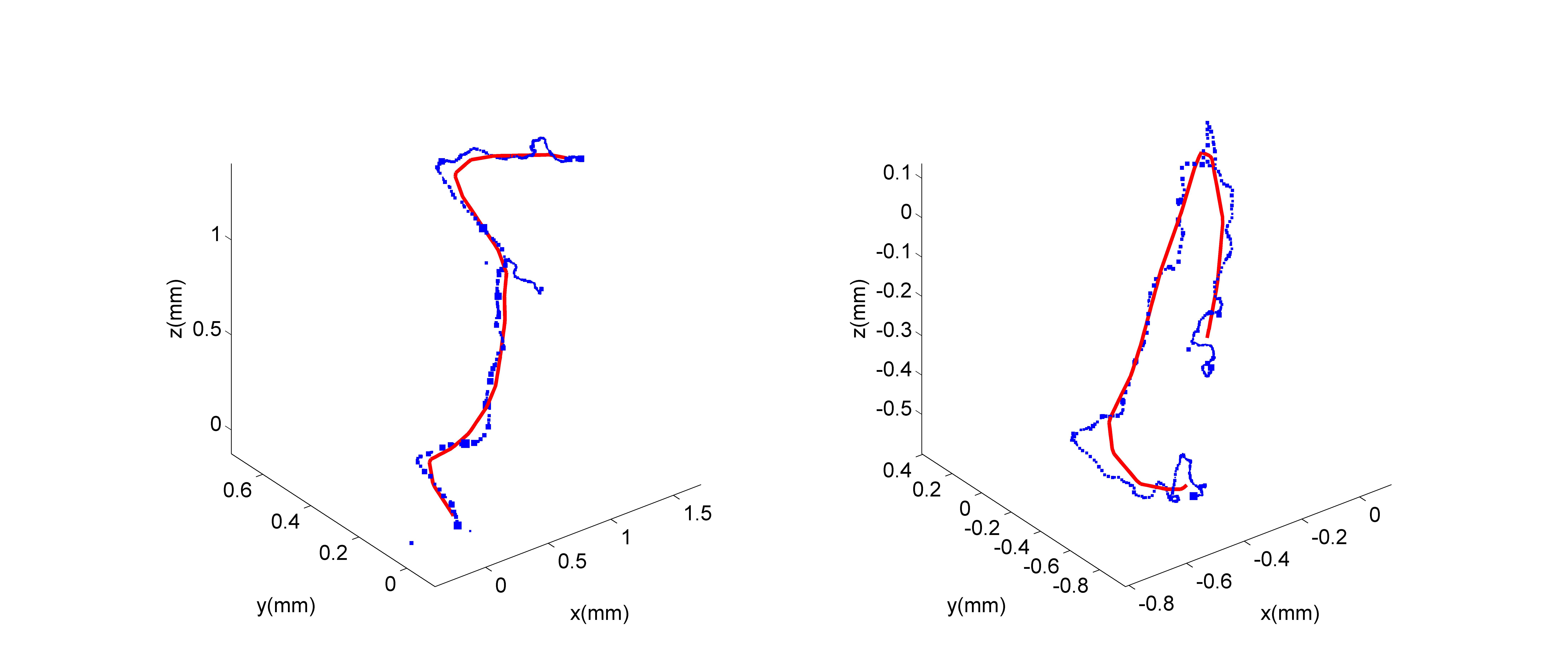}
    \caption{\label{fig8} An instance of the comparison between the MC track and the \textit{REF track} of a single-electron event (left) and a 0$\nu$$\beta$$\beta$ event (right). The differences are mainly caused by the information loss during voxelization.}

    \includegraphics[width=15cm]{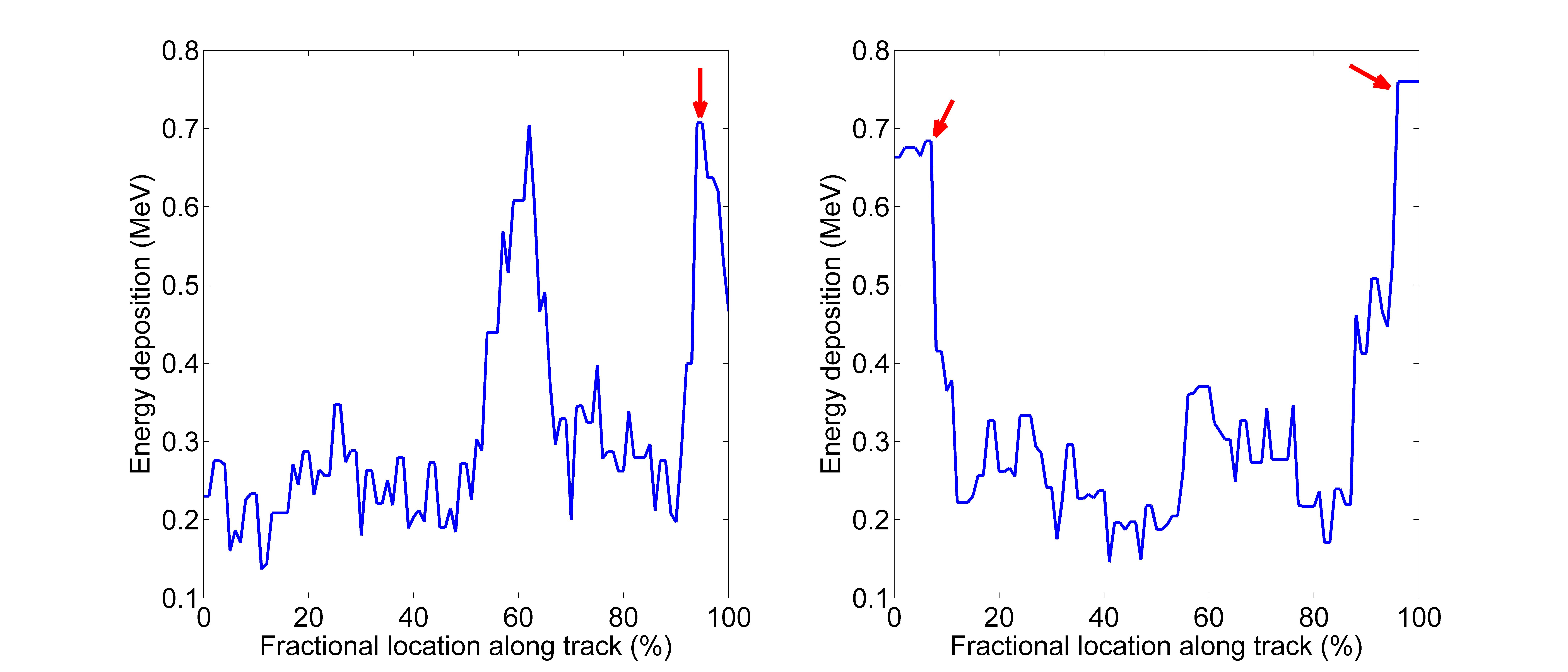}
    \caption{\label{fig9} An instance of the energy deposition along the \textit{REF track} of the single-electron event (left) and a 0$\nu$$\beta$$\beta$ event (right) in Fig.~\ref{fig7}. The topological signature of ¡®blobs¡¯ can be found as 1 peak (left) or 2 peaks (right) near the ends of the curve.}
    \end{center}
\end{figure}

Fig.~\ref{fig8} shows the comparison between the \textit{REF track} and the MC track, which proves that the \textit{REF track} is quite an accurate approximation of the real track of the particle and can be qualified as a crucial topological signature to study the track features of electrons.

\subsection{Calculation of the \textit{REF energy deposition}}
{The new topological signature \textit{REF track} mainly provides the position information of the track, and then the energy features should be considered.} With the \textit{REF track} of single-electron and 0$\nu$$\beta$$\beta$ events, the energy deposition in an \textit{R}-radius ball along the \textit{REF track} is calculated, which is called \textit{REF energy deposition}. At every point \textit{J}$_{k}$ on the \textit{REF track}, the energy deposition \textit{E}$_{k}$ is defined as the sum of the energy in the \textit{R}-radius ball around \textit{J}$_{k}$.

As the \textit{REF track} of the same type of events may also vary in length, \textit{J}$_{k}$ is chosen at the 0\%, 1\% $\cdots$ 100\% factional lengths of the \textit{REF track} (\textit{J}$_{0}$ at 0\% and \textit{J}$_{100}$ at 100\% are the two endpoints). Since the \textit{REF track} has no direction, 0\% is defined as the lower-energy-deposition endpoint of the \textit{REF track} and 100\% is defined as the other endpoint (i.e., \textit{E}$_{0}$ $<$ \textit{E}$_{100}$). For every event, the values of the \textit{REF energy deposition} \textit{E}$_{0}$, \textit{E}$_{1}$, \textit{E}$_{2}$ $\cdots$ \textit{E}$_{100}$ can be calculated.

An instance of the \textit{REF energy deposition} is shown in Fig.~\ref{fig9}. The topological signature of the `blobs' can be found as peaks near the two ends of the curve.

\section{Discrimination methods and results}

\subsection{\textit{REF energy deposition} model}
With the Monte Carlo simulation method described in section 2, 10$^{5}$  single-electron events and 10$^{5}$  0$\nu$$\beta$$\beta$ events are generated. For each event after pre-selection, the \textit{REF energy deposition} \textit{E}$_{0}$, \textit{E}$_{1}$, \textit{E}$_{2}$ $\cdots$ \textit{E}$_{100}$ at each factional length \textit{J}$_{0}$, \textit{J}$_{1}$, \textit{J}$_{2}$ $\cdots$ \textit{J}$_{100}$ are calculated.

\begin{figure}[!htb]
    \begin{center}
    \includegraphics[width=7.5cm]{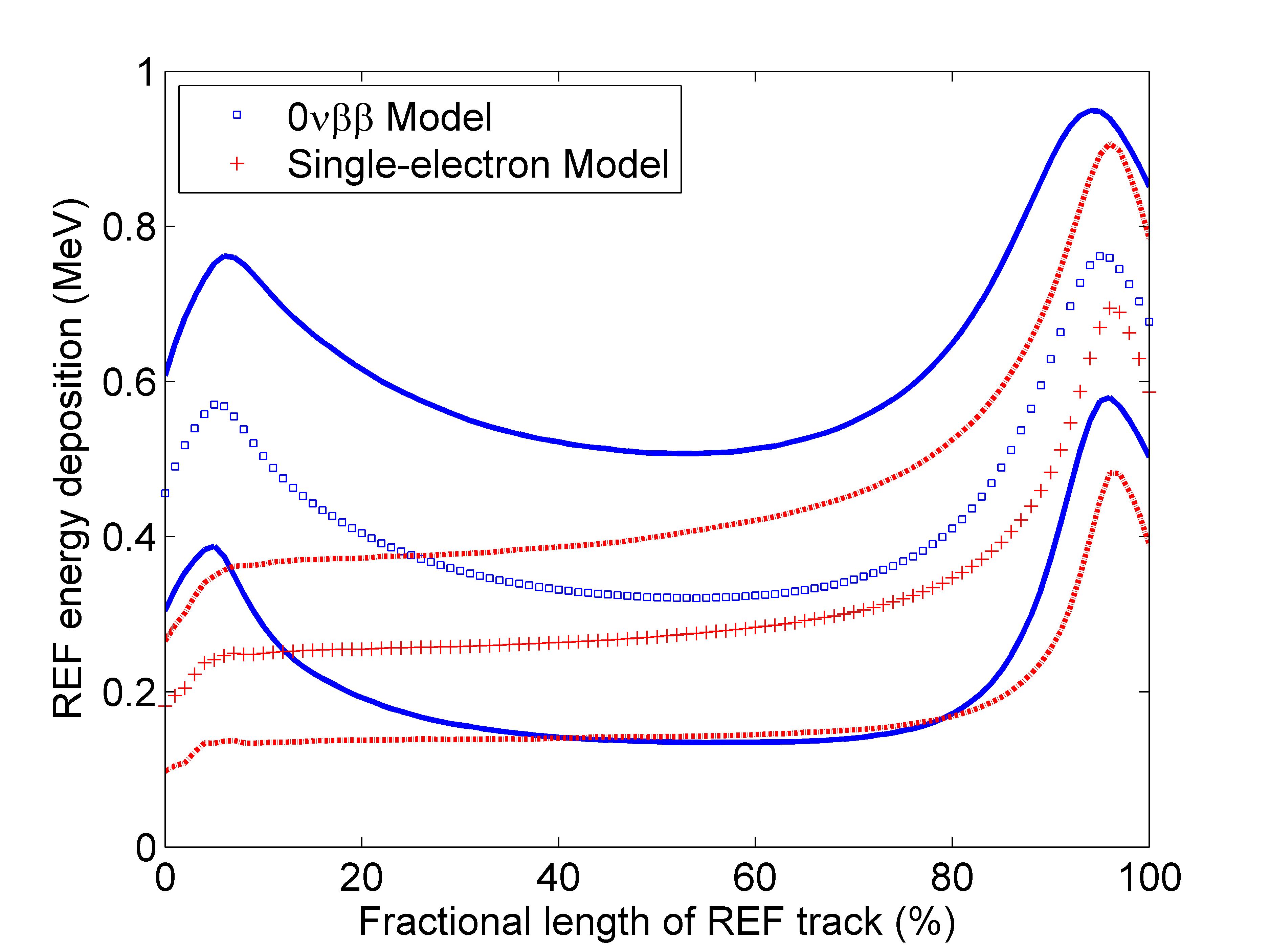}
    \caption{\label{fig10} The \textit{REF energy deposition models} of the single-electron events and 0$\nu$$\beta$$\beta$ events. {The square (plus sign) markers indicate $\overline{\textit{E}_{k}}$ of 0$\nu$$\beta$$\beta$ events (single-electron events). The lines above and below the markers indicate $\sigma$$_{k}$ of each type of events. The features of `one blob' and `two blobs' can also be clearly found in the model. There are large deviations of energy depositions in the middle of the \textit{REF track}, reflecting the high fluctuation of energy depositions due to multiple scattering.}}
    \end{center}
\end{figure}

For all of the single-electron events generated, the sample mean value $\overline{\textit{E}_{0}}$, $\overline{\textit{E}_{1}}$, $\overline{\textit{E}_{2}}$ $\cdots$ $\overline{\textit{E}_{100}}$ and the sample standard deviation $\sigma$$_{0}$, $\sigma$$_{1}$, $\sigma$$_{2}$ $\cdots$ $\sigma$$_{100}$ can be calculated at each \textit{J}$_{k}$. The value of $\overline{\textit{E}_{k}}$ reflects the average energy deposition at \textit{J}$_{k}$, and the value of $\sigma$$_{k}$ reflects the fluctuation of the energy deposition at \textit{J}$_{k}$. {The $\overline{\textit{E}_{k}}$ and $\sigma$$_{k}$ features of the single-electron events are defined as a \textit{REF energy deposition model} of them. Similarly, the \textit{REF energy deposition model} of 0$\nu$$\beta$$\beta$ events can be calculated as well. Both models are shown in Fig.~\ref{fig10}.}

Fig.~\ref{fig10} indicates that there {are significant differences} between the \textit{REF energy deposition models} of single-electron events and 0$\nu$$\beta$$\beta$ events. The apparent differences between the two models make it possible to develop a new discrimination method.

\subsection{Discrimination methods}
The \textit{REF energy deposition models} show the difference between the single-electron events and the 0$\nu$$\beta$$\beta$ events, which can be used to identify whether an unknown track is a single-electron event or a 0$\nu$$\beta$$\beta$ event.

To describe the features of the \textit{REF energy deposition} of an unknown track, a vector is defined:
\begin{eqnarray}
    \label{eq2}
    \vec E =  (E_{0},E_{1},E_{2}, \cdots ,E_{100})
\end{eqnarray}

In the view of \textit{pattern recognition}, the discrimination between single-electron events and 0$\nu$$\beta$$\beta$ events involves finding a separation between $\vec{\textit{E}}$ of single-electron events and $\vec{\textit{E}}$ of 0$\nu$$\beta$$\beta$ events. One approach to obtain the maximum separation is the Fisher's linear discriminant (see, for example, Ref. \cite{lab18}).

A linear combination of $\vec{\textit{E}}$ is established:
\begin{equation}
    \label{eq3}
    S = \vec w \cdot \vec E={\sum_{k=0}^{100} w_k \cdot E_k}
\end{equation}

where
\begin{equation}
    \label{eq6}
        \vec w= (w_{0},w_{1},w_{2}, \cdots ,w_{100})
\end{equation}

\textit{w}$_{k}$ is the weighting factor for \textit{E}$_{k}$, which indicates the significance of \textit{E}$_{k}$ at the fractional distance \textit{k}. The evaluation of \textit{w}$_{k}$ is crucial to the discrimination method and can be evaluated in the Fisher's linear discriminant as (the $\vec{\textit{E}}$ of single-electron events have means $\vec{\textit{$\mu$}}_{Single}$ and covariance {matrix} ${\textit{$\Sigma$}_{Single}}$, and the $\vec{\textit{E}}$ of 0$\nu$$\beta$$\beta$ events have means $\vec{\textit{$\mu$}}_{Double}$ and covariance {matrix} ${\textit{$\Sigma$}_{Double}}$):
\begin{eqnarray}
    \label{eq4}
    \vec w = (\Sigma_{Double}+\Sigma_{Single})^{-1}(\vec{\mu}_{Double}-\vec{\mu}_{Single})
\end{eqnarray}

The cut \textit{c} in Fisher's linear discriminant can be evaluated as:
\begin{eqnarray}
    \label{eq5}
    c = \vec w \cdot \frac{1}{2}(\vec{\mu}_{Double}+\vec{\mu}_{Single})
\end{eqnarray}

If \textit{S} $>$ \textit{c}, the unknown track is identified as a 0$\nu$$\beta$$\beta$ event; otherwise, for \textit{S} $<$ \textit{c}, the track is identified as a single-electron event.

\begin{figure}[!htb]
    \begin{center}
    \includegraphics[width=7.5cm]{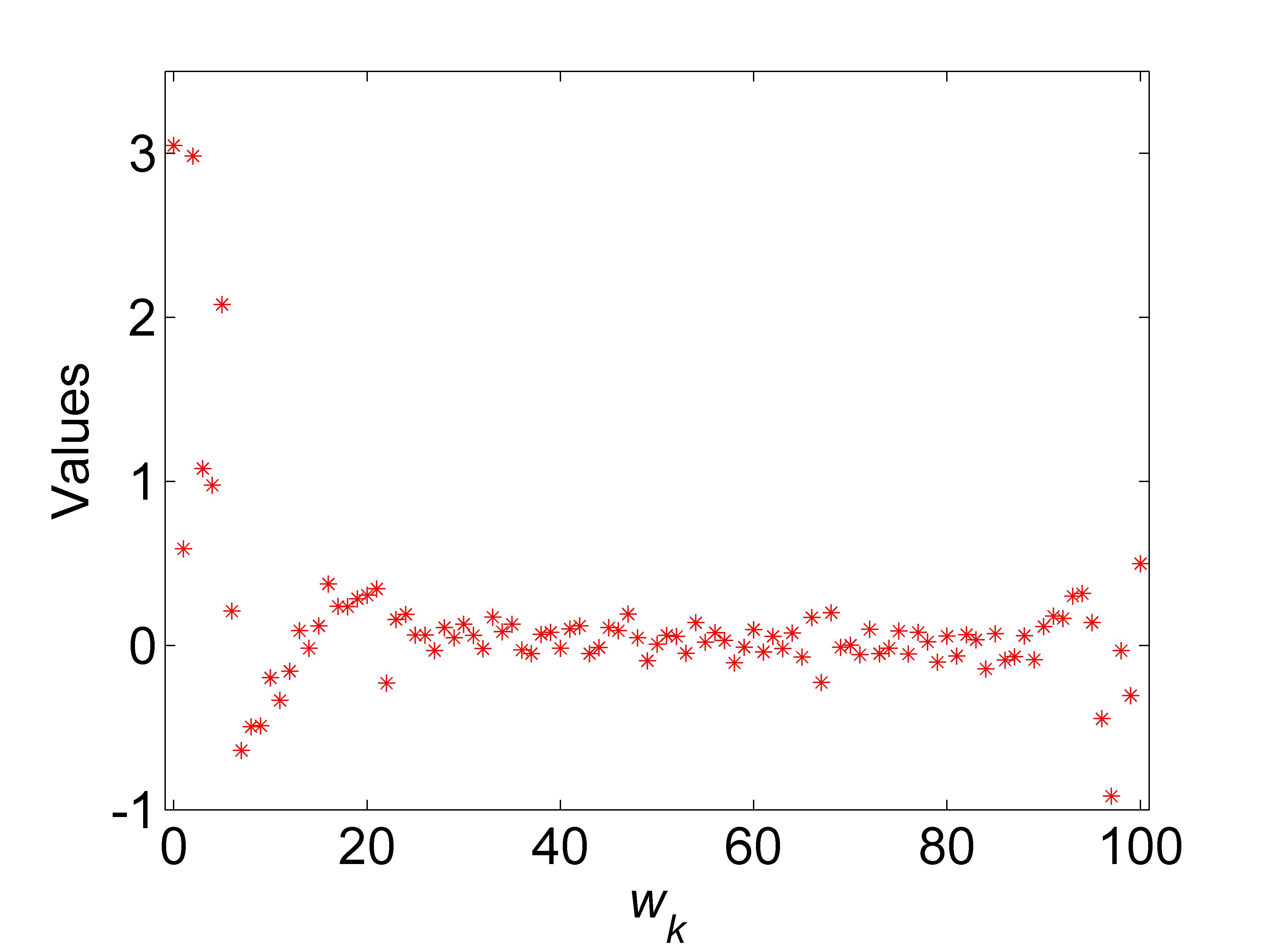}
    \caption{\label{fig11} {An example of the \textit{w}$_{k}$ values. w$_{0}$ is the largest one. \textit{w}$_{k}$ can be negative because the covariance matrix ${\textit{$\Sigma$}_{Single}}$ and ${\textit{$\Sigma$}_{Double}}$ may have negative elements.}}
    \end{center}
\end{figure}

Note that different values of \textit{D} (side length of voxel), \textit{R} (radius of energy-deposition ball), and \textit{T} (\textit{connectivity criterion}) will result in different models, thus producing different \textit{w}$_{k}$ and \textit{c}. However, in general, w$_{0}$ is always the largest one of all of the \textit{w}$_{k}$, i.e., the most significant variance between the 0$\nu$$\beta$$\beta$ model and the single-electron model is at the endpoint of \textit{k} = 0 (See Fig.~\ref{fig11}). When only using the single value of the energy deposition at the lower-energy-endpoint, the method is simplified and equivalent to the method of the NEXT experiment \cite{lab9}.

\subsection{Discrimination efficiency}
For every single-electron event and 0$\nu$$\beta$$\beta$ event, the value of \textit{S} is calculated and compared with \textit{c} {to categorize the event}. By comparing the identified results with the real results, the discrimination efficiency of both single-electron events and 0$\nu$$\beta$$\beta$ events can be calculated.

The discrimination efficiency is influenced by the choice of \textit{D}, \textit{R}, and \textit{T}. In this section, the value is set as \textit{D} = 0.1 mm, \textit{R} = $\sqrt{3}$\textit{D}, and \textit{T} = $\sqrt{3}$\textit{D} by default.

The efficiency of the discrimination method is shown in Table~\ref{tab1}. The single-electron background is rejected with a factor of 93.8 $\pm$ 0.3 (stat.)\%, with an efficiency of 85.6 $\pm$ 0.4 (stat.)\% for the 0$\nu$$\beta$$\beta$ signal events. {The statistical uncertainties were obtained by repeating the method in different Monte Carlo datasets and calculating the standard deviation of efficiencies. The systematic uncertainties were not considered in this paper and would be discussed in future works.}
\begin{table}[h]
    \begin{center}
    \caption{ \label{tab1}  Efficiency of the discrimination method.}
    \footnotesize
    \begin{tabular*}{80mm}{c@{\extracolsep{\fill}}cc}
    \toprule \multirow{2}{*}{Conditions} &\multicolumn{2}{c}{\textit{D} = 0.1 mm, \textit{R} = $\sqrt{3}$\textit{{D}}, \textit{T} = $\sqrt{3}$\textit{{D}}}     \\
     & Single-electron events&0$\nu$$\beta$$\beta$ events   \\
    \hline
    Efficiency & 93.8 $\pm$ 0.3 (stat.)\% & 85.6 $\pm$ 0.4 (stat.)\% \\
    \bottomrule
    \end{tabular*}
    \vspace{0mm}
    \end{center}
    \vspace{0mm}
\end{table}

\section{Discussions}
\subsection{The effects of \textit{T} (\textit{connection criterion})}
The \textit{connection criterion} \textit{T} influences the {shape} of the \textit{primary path}. Only when the space distance between two \textit{points} is no greater than \textit{T} are the \textit{points} connected with a \textit{line}. Different values of \textit{T} have different implications in the \textit{graph}:

\textit{T} = \textit{D}: Two \textit{points} in the \textit{graph} are connected if their voxels share a face.

\textit{T} = $\sqrt{2}$\textit{D}: Two \textit{points} are connected if their voxels share a face or an edge.

\textit{T} = $\sqrt{3}$\textit{D}: Two \textit{points} are connected if their voxels share a face, an edge or a corner.

Among these cases, the smallest \textit{T} = \textit{D} is the most strict criterion, and in this case, the distance between every neighbour point in the \textit{primary path} must equal \textit{D}. For comparison, for the case of \textit{T} = $\sqrt{3}$\textit{D}, this distance can be \textit{D}, $\sqrt{2}$\textit{D}, or $\sqrt{3}$\textit{D}.

In general, the \textit{T} = \textit{D} criterion retains the most detailed information because, in this case, the \textit{primary path} would pass through the greatest number of voxelized points. As a result, the \textit{REF track} reconstructed from the \textit{T} = \textit{D} \textit{primary path} best mimics the MC track.

However, if the energy resolution of the detector and electronic noise are taken into account, the performance of the \textit{T} = \textit{D} criterion will degrade. For example, if the energy deposition of a \textit{point} X in the \textit{primary path} decreases to zero due to the {noise}, the \textit{primary path} algorithm will probably return two shorter paths instead of the original one, thus resulting in an incorrect \textit{primary path}. {Hence, the \textit{connection criterion} \textit{T} was chosen as $\sqrt{3}$\textit{D} to minimize the influence of isolated breakpoints resulting from limited energy resolution or electronic noise.}

\subsection{The effects of \textit{R} (radius of the energy-deposition ball)}
For each event, the value of \textit{R} influences the effects of the spatial energy filter, thus resulting in different \textit{REF track}s. The filter is effective only when \textit{R} is greater than \textit{D}. Otherwise, if \textit{R} $<$ \textit{D}, only one voxelized point can be found in the energy-deposition ball; as a result, the filter is disabled and the \textit{REF track} and the \textit{primary path} are identical. Meanwhile, \textit{R} should not be too large because the detailed information of the track will be lost.

This section takes \textit{D} = 0.1 mm and \textit{T} = $\sqrt{3}$\textit{D} as the default condition and discusses how the variation of \textit{R} influences the discrimination efficiency. {The results are calculated from a same Monte Carlo dataset and are listed in Table~\ref{tab2}.}

\begin{table}[h]
    \begin{center}
    \caption{ \label{tab2}  Efficiency of the discrimination method with different \textit{R}.}
    \footnotesize
    \begin{tabular*}{170mm}{@{\extracolsep{\fill}}ccccccccc}
    \toprule Conditions & \textit{R} $<$  \textit{D}  & \textit{R} = \textit{{D}} & \textit{R} = $\sqrt{2}$\textit{{D}}&  \textit{R} = $\sqrt{3}$\textit{{D}}& \textit{R} = 2\textit{{D}}  &  \textit{R} = $\sqrt{5}$\textit{{D}} &  \textit{R} = $\sqrt{6}$\textit{{D}} &  \textit{R} = $\sqrt{7}$\textit{{D}} \\
    \hline
    Single-electron rejection & 83.1\% & 90.6\% & 93.1\% & 93.8\% & 92.7\% & 92.1\% & 91.7\% & 91.4\%\\
    0$\nu$$\beta$$\beta$ efficiency & 70.3\% & 74.9\% & 82.4\% & 85.6\% & 86.1\% & 86.8\% & 87.8\% & 88.0\%\\
    \bottomrule
    \end{tabular*}%
    \end{center}
\end{table}

Table~\ref{tab2} shows that the discrimination efficiency of an \textit{R} $<$ \textit{D} filter (disabled filter, \textit{REF track} = \textit{primary path}) is much lower than that of an \textit{R} = $\sqrt{3}$\textit{D} filter{, which shows the advantage of  \textit{REF track} over \textit{primary path}.

To explain this advantage quantitatively, the reconstructed \textit{REF track} ( \textit{primary path}) is compared with their MC track by calculating the distance between each point in the MC track and the nearest point in the \textit{REF track} ( \textit{primary path}). Then the proportion of large distances among total distances is recorded. The proportion of distance over 0.1 mm is in average 20\% for the \textit{REF track} and 27\% for the \textit{primary path}, and the proportion of distance over 0.2 mm is in average 6\% for the \textit{REF track} and 9\% for the \textit{primary path}. It proves that the \textit{REF track} always meets less deviation than the \textit{primary path} when compared with MC track. Hence, the \textit{REF track} introduced in this paper proves a better approximation of the MC tracks (i.e. real tracks) than the \textit{primary path}, and results in a better performance in terms of the discrimination efficiency.
}

In addition, Table~\ref{tab2} shows the best filter for a high efficiency for both single-electron events and 0$\nu$$\beta$$\beta$ events can be found at approximately \textit{R} = $\sqrt{3}$\textit{D}.

\subsection{The effects of \textit{D} (side length of the voxel)}
According to the scales of the length of the electron's track (approximately 3.7 mm at 2813.50 keV) described in section 3.1.1, the 3-D spatial resolution of a track should be as good as several submillimetres to obtain the topological signatures of an electron's track. In this paper, the side length of the voxel is set to a default of 0.1 mm (\textit{D} = 0.1 mm) to denote a spatial resolution of 0.1 mm.

If the spatial resolution is worse than 0.1 mm, then the performance of this discrimination method begins to degrade in terms of the discrimination efficiency of both single-electron events and 0$\nu$$\beta$$\beta$ events. This paper evaluates this degradation by setting a larger value of \textit{D}, corresponding to the spatial resolution of a practical detector.

{The results of the discrimination efficiency affected by \textit{D} are shown in Table~\ref{tab3} and they are calculated from the same Monte Carlo dataset}. For different \textit{D} values, the value of \textit{R} should be reappraised. Only the optimal \textit{R} is listed in the table.

\begin{table}[h]
    \begin{center}
    \caption{ \label{tab3}  Efficiency of the discrimination method with different \textit{D} and optimal \textit{R}.}
    \footnotesize
    \begin{tabular*}{170mm}{@{\extracolsep{\fill}}cccccc}
    \toprule \multirow{2}{*}{Conditions} & \textit{D} = 0.1 mm & \textit{D} = 0.2 mm & \textit{D} = 0.3 mm & \textit{D} = 0.4 mm & \textit{D} = 0.5 mm \\
     & \textit{R} = $\sqrt{3}$\textit{{D}}  & \textit{R} = $\sqrt{2}$\textit{{D}} & \textit{R} = $\sqrt{2}$\textit{{D}}&  \textit{R} = \textit{{D}}& \textit{R} = \textit{{D}}   \\
    \hline
    Single-electron rejection & 93.8\% & 85.9\% & 84.8\% & 83.5\% & 81.5\% \\
    0$\nu$$\beta$$\beta$ efficiency &85.6\% & 81.4\% & 79.5\% & 71.7\% & 70.7\% \\
    \bottomrule
    \end{tabular*}%
    \end{center}
\end{table}

As \textit{D} increases, the loss of the {track} information becomes much more serious; as a result, \textit{R}/\textit{D} in the optimal performance should be smaller to reduce further information loss. When \textit{D} is 0.4 mm or 0.5 mm, \textit{R} = \textit{D} provides the optimal \textit{R} with an effective filter. (Even in these cases, \textit{R} $<$ \textit{D} filter performs worse than \textit{R} = \textit{D} filter.)
From Table~\ref{tab3}, it can be concluded that even if the spatial resolution of a CdZnTe detector degrades to 0.5 mm, this discrimination method can still provide {a rejection factor of 82\% for single-electron events and an efficiency of 71\% for 0$\nu$$\beta$$\beta$ events.}

\subsection{The effect of \textit{c} (cut in Eq. 6)}

{

In the discrimination method described in section 4.3, the cut \textit{c} is evaluated as Eq. 6. Then the value of \textit{S}  for each event is calculated and compared with \textit{c}, thus obtaining the identified result. Fig.~\ref{fig10} shows the distributions of the values of \textit{S} for single-electron events and 0$\nu$$\beta$$\beta$ events. The distributions are almost Gaussian, and the cut \textit{c} in Eq. 6 is chosen as the center of mean values of two Gaussian distributions. The ratio of \textit{S} below (above) \textit{c} in single-electron events (0$\nu$$\beta$$\beta$ events) indicates the discrimination efficiency for single-electron events (0$\nu$$\beta$$\beta$ events).

\begin{figure}[!htb]
    \begin{center}
    \includegraphics[width=7.5cm]{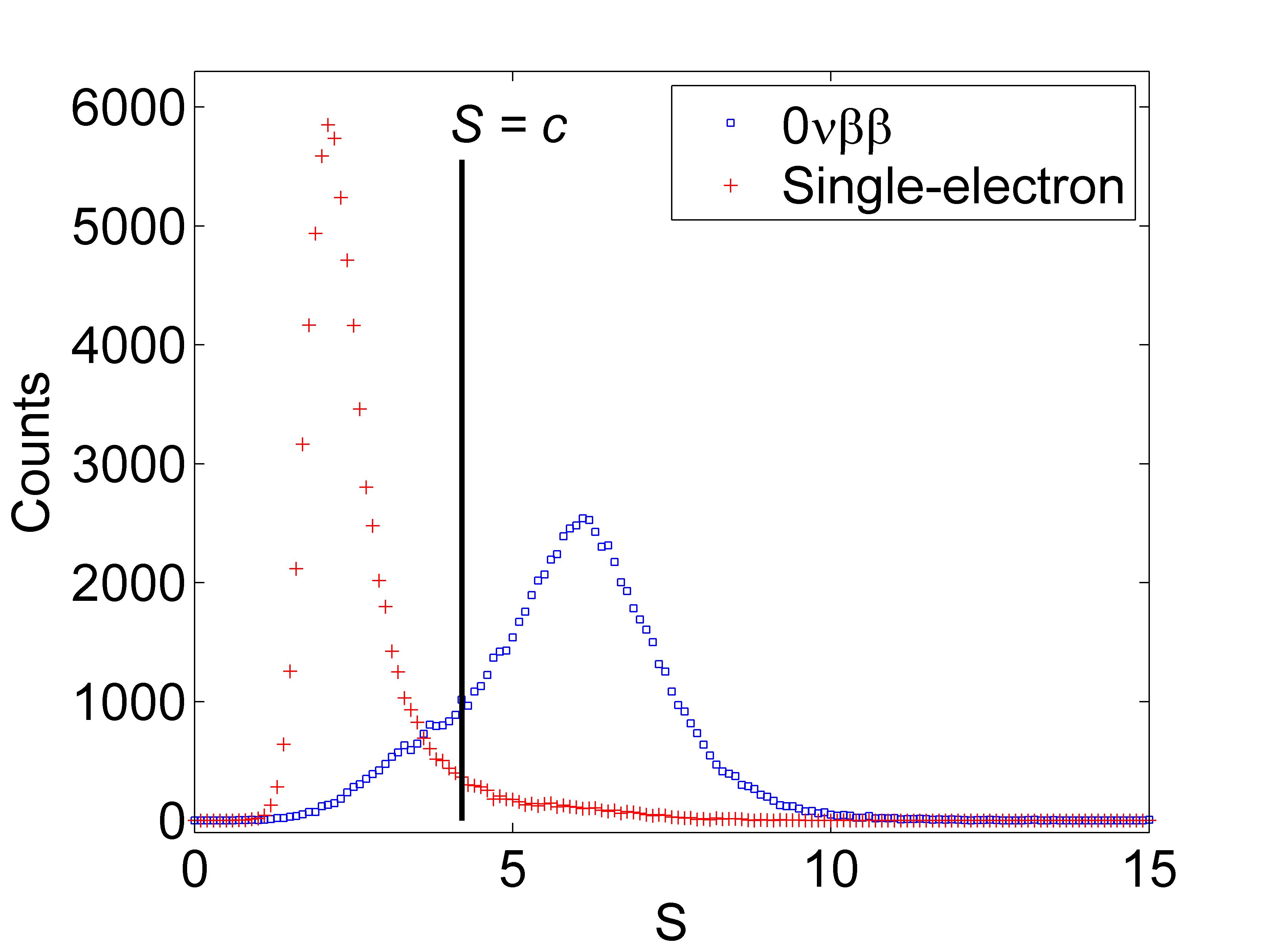}
    \caption{\label{fig12} The distributions of the values of \textit{S} for single-electron events and 0$\nu$$\beta$$\beta$ events.}
    \end{center}
\end{figure}

It is possible to increase the value \textit{c} to obtain higher single-electron rejection factor at the cost of a lower 0$\nu$$\beta$$\beta$ efficiency, which may optimize the figure of merit \textit{F = s /$\sqrt{b}$} (\textit{s} is the 0$\nu$$\beta$$\beta$ efficiency and \textit{b} is fraction of single-electron events that survives the requirements for 0$\nu$$\beta$$\beta$ signals). For optimal figures of merit, this discrimination method would have 0$\nu$$\beta$$\beta$ efficiency of 70.7\% and single-electron rejection factor of 96.7\%.
}

\subsection{The effect of the pre-selection in the MC simulation}
{
In section 2.1, a pre-selection is made to exclude the energy-not-fully-deposited events, which are mainly caused by bremsstrahlung photons' escape. When the size of the CdZnTe crystal is 10 mm $\times$ 10 mm $\times$ 10 mm, about 74\% single-electron and 81\% 0$\nu$$\beta$$\beta$ events deposit all of their energy within the crystal boundaries. For the events which pass the pre-selection, a discrimination efficiency of 93.8\% (85.6\%) for single-electron (0$\nu$$\beta$$\beta$) is obtained.

For comparison, the discrimination method was also applied to all events before pre-selection, and a discrimination efficiency of 93.8\% (84.6\%) for single-electron (0$\nu$$\beta$$\beta$) events was obtained, which didn't meet an apparent degradation. The reason is that the \textit{REF energy deposition model} (Fig.~\ref{fig10}) focuses on the energy features along an electron's (two electrons') track(s), and is not so sensitive to the electron's energy. When a bremsstrahlung photon escapes, the electron tracks left in the crystal may still carry the energy features for identifying it as a single-electron or 0$\nu$$\beta$$\beta$ event.

This inspiring result provides a possibility to discriminate single-electron and 0$\nu$$\beta$$\beta$ events in large energy intervals, which will be discussed in details in the future.


}
\subsection{Energy resolution and electronics noise}

This paper aims to propose a methodology for discussing the topological signatures and potential discrimination methods for single-electron and 0$\nu$$\beta$$\beta$ events, without considering the detector energy resolution, electronics noise and other effects in a practical experimental implementation which should be a dedicated detector system. Because of the current technical limitations in achieving a high 3-D spatial resolution in large volume CdZnTe, the feasible experimental approach to realize a 3-D track reconstruction in CdZnTe detectors is still difficult and requires further research. {Nevertheless, with progress on hybrid pixel detectors and reconstruction methods developed by Mykhaylo Filipenko \cite{lab13} and He Zhong \cite{lab19}\cite{lab20}, obtaining a 3-D track of about 0.1 mm spatial resolution in a CdZnTe of about 10$^{3}$ mm$^{3}$ is promising in the near future. In Ref. \cite{lab13}, under several assumptions, about 0.1 mm 3-D spatial resolution has already been achieved in a 14 mm $\times$ 14 mm $\times$ 1 mm CdTe-Timepix system.}

\section{Conclusion}

In this paper, a new topological signature `\textit{REF track}' of single-electron events and 0$\nu$$\beta$$\beta$ events was introduced in this paper and proved to be {a good approximation} of their real tracks. {Derived from} the energy deposition along the \textit{REF track}, we propose a second topological signature `\textit{REF energy deposition}' to {characterize} the energy features of both events.

Based on the analysis of large amounts of single-electron events and 0$\nu$$\beta$$\beta$ events in Monte Carlo simulation, this paper established the \textit{REF energy deposition model} {characterizing the differences between two kinds of events}, and proposed a new discrimination method applying Fisher's linear discriminant. With a criteria of \textit{D} = 0.1 mm, \textit{R} = $\sqrt{3}$\textit{D}, and \textit{T} = $\sqrt{3}$\textit{D}, a performance with a signal efficiency of  85.6 $\pm$ 0.4 (stat.)\% and a single-electron background rejection of 93.8 $\pm$ 0.3 (stat.)\% was {achieved}. This discrimination method has features of both a high robustness and high discrimination efficiency compared with the existing lower-energy single blob method{, since} this method considers all of the energy features along the entire track. Although this discrimination method is designed for distinguishing single-electron events and 0$\nu$$\beta$$\beta$ events, it is also possible to apply it for rejecting short tracks of alpha rays and straight tracks of muons, making this method feasible for application in future research.

This paper also discussed how the values of \textit{D}, \textit{R}, \textit{T} and \textit{c} affect the discrimination efficiency and proposed a possibility to discriminate single-electron and 0$\nu$$\beta$$\beta$ events of different energies. Although due to current technical limitations conducting a practical experiment is difficult for {the} moment, recent progress on hybrid pixel detector technologies and reconstruction methods \cite{lab13}\cite{lab19}\cite{lab20}makes it promising to propose an experiment in the very near future, especially with Timepix3 when it is commercially available.

\section*{References}

\bibliography{mybibfile}

\end{document}